\documentclass{prop2015}% no class options needed by now
\usepackage[english]{babel}
\title{Soft Theorems from String Theory}
\author[P. Di Vecchia]{P. Di Vecchia\inst{1,2}}
\author[R.\,Marotta]{R. Marotta\inst{3}}
\author[M.\,Mojaza]{M. Mojaza\inst{2}}
\address[1]{ The Niels Bohr Institute, University of Copenhagen, Blegdamsvej 17, 
DK-2100 Copenhagen {\O}, Denmark.}
\address[2]{Nordita, KTH Royal Institute of Technology and Stockholm University, Roslagstullsbacken 23, SE-10691 Stockholm.}
\address[3]{Istituto Nazionale di Fisica Nucleare, Sezione di Napoli, Complesso
 Universitario di Monte S. Angelo ed. 6, via Cintia, 80126, Napoli, Italy.}
\shortauthors{P. Divecchia et al.}
\begin{abstract}
Soft behaviour of closed string amplitudes involving  dilatons, gravitons and 
anti-symmetric tensors, is  studied in the framework of bosonic string theory. 
The leading double soft limit of gluons is analysed as well, starting from 
scattering amplitudes computed in the open bosonic string. Field theory 
expressions  are then obtained by sending the string tension to infinity. 
The  presented  results  have been  derived in the papers of Ref~\cite{1502.05258}. 
\end{abstract}
\shortabstract
\begin{document}
\maketitle
%%% Use this if the article text won't start with a \section:
% \noindent
%%% Being based on LaTeX's article class, and2012 supports the respective 
%%% sectioning level from \section to \subparagraph.

\section{Introduction}

There is a long history of studying the relations between the soft
%infrared
 behaviour 
of field theory amplitudes and the symmetries of the underlying theory.

Since the 50s  it has been shown that the  leading behaviour of  scattering  
amplitudes with a soft photon is obtained by means of gauge invariance  
from  the  corresponding amplitudes  without the soft particle\cite{LowFourPt}. 
The extension of this theorem to the universal leading behaviour of amplitudes 
with one soft graviton was discussed by Weinberg in the 60s\cite{Weinberg}.

The nonlinear realization of symmetries provides another example of the 
relation existing  between  symmetries and low energy theorems. When a 
group $G$ is spontaneously broken to some subgroup $H$, Nambu-Goldstone 
bosons appear and  parametrize the coset space $G/H$. Their interaction with 
the other fields charged under the symmetry group is  described in the Lagrangian 
 by  derivative couplings. As a result, amplitudes involving one soft 
Nambu-Goldstone boson are vanishing\cite{ArkaniHamed:2008gz,1412.2145}. 
These are the famous Adler's zero studied for the first time in the contest of 
pion dynamics\cite{adler}. 

Recently the leading divergent behaviour of amplitudes with a soft graviton was 
obtained from the Ward identity\cite{Strominger} of the diagonal Bondi, van der 
Burg, Metzner and Sachs supertranslation symmetry\cite{BMS}. Later, similar 
results have been obtained  for Yang-Mills theory where the soft gluon theorem 
arises as the Ward identity of a two dimensional Kac-Moody type 
symmetry\cite{mitra}.  
The extension of these theorems to subleading order for gluons, and 
sub-subleding order for gravitons, have been obtained by computing 
on-shell scattering amplitudes\cite{all,Bianchi2015} and proved  in 
arbitrary dimensions by using Poincar\'e and on-shell gauge 
invariance\cite{gauge,BDDN}. 
In these new soft theorems, $n+1$-point amplitudes with a soft graviton or 
gluon  are obtained acting on $n$-point hard amplitudes with universal  soft 
operators depending on the momenta and polarizations of the hard particles. 

The symmetries of the quantum field theory are also reflected in the double 
soft behaviour of scattering amplitudes with scalar particles or gluons. This has 
been made explicit in Ref.\cite{ArkaniHamed:2008gz} where, in the case of 
spontaneously broken symmetries, it has been shown that  amplitudes with 
two soft Nambu-Goldstone bosons capture the algebra of the broken 
generators of the global symmetry.

More recently supergravity amplitudes involving fermion particles have been 
studied in three and four dimensions and  in the kinematic region where two 
of these particles carry small momentum\cite{1412.1809}. Subleading terms 
in the emission of two soft scalars computed in Cachazo-He-Yuan (CHY) 
representation of the amplitude, have been determined for a vast class of  
theories in Ref.\cite{Cachazo:2015ksa}. In the spinor helicity formalism, similar 
analyses are performed for amplitudes with  gravitons and gluons and for 
scalars of  ${\cal N}=4$ super-Yang-Mills in Ref.\cite{Klose:2015xoa}. New 
soft theorems in gauge theories with more than one particle  are 
derived in \cite{Volovich:2015yoa} both in four and in any dimensions 
by using respectively the BCFW and CHY formula.

In this very short paper, which is a summary of the main results obtained 
in Ref.\cite{1502.05258}, a purely string approach to the low energy 
theorems is presented. 

Soft gluon and graviton behaviour was also studied in the framework 
of string theory in 
Refs.\cite{Ademollo:1975pf,StringSoft,Volovich:2015yoa,1505.05854}. 
In these papers it has been shown that string amplitudes reproduce the 
soft theorems without any $\alpha'$ correction, $\alpha'$ being the string 
slope\cite{Bianchi2015}.

In this work we consider bosonic string theory and do not only  confirm 
the results obtained in the literature regarding the emission of  soft 
gravitons, but also extend them to the dilaton and Kalb-Ramond fields. 
In the case of the Kalb-Ramond we do not get any pole term and we  find a  
peculiar relation between $n+1$ point amplitudes with  a soft antisymmetric 
tensor and $n$ point hard amplitudes, which involve, as an intermediate 
step, the introduction of holomorphic and antiholomorphic momenta. 
This handling  of the momenta is quite natural in closed string theory, but 
the relation obtained between  amplitudes with and without the soft particle  
is not a real low energy theorem, because the hard amplitudes are not physical. 

The amplitude with a soft graviton and $n$ tachyons is obtained through  
second subleading order. According to a standard trick, the tachyon is 
seen as a scalar field with mass $m^2=-4/\alpha'$ and therefore we 
have an example of validity  of the  low energy theorem for massive matter. 

String theory is also a powerful tool to get field theory amplitudes. There are 
few diagrams at each order of the perturbative expansion that  are represented 
as complex integrals on the string  moduli space.   
We have used this compact representation of scattering amplitudes to 
compute, in  bosonic string theory, the colour ordered amplitude with 
$n+2$ gluons. On this amplitude two different  double soft limits are  
performed. In one case, contiguous gluons are taken with small 
momentum, in the other case two soft gluons are separated by a hard particle. 
In both examples gauge invariant expressions are derived.

\section{Single soft limit of string amplitudes}
\label{tachyon}
\setcounter{equation}{0}

The scattering amplitude involving a massless closed string state, graviton or dilaton, and $n$ closed string tachyons is given by the tensor:
\begin{eqnarray}
M_{\mu \nu}= \frac{8\pi}{\alpha'}\bigg(\frac{\kappa_d}{2\pi}\Bigg)^{n-1}\int \frac{\prod_{i=1}^{n} d^2 z_i}{d V_{abc}} \prod_{i<j} |z_i - z_j|^{\alpha' k_i k_j} 
S _{\mu \nu} \, ,
\label{M1n} 
\end{eqnarray}
where
\begin{eqnarray}
S_{\mu \nu}  =  \frac{\alpha'}{2}  \int d^2 z \prod_{\ell=1}^{n} | z- z_{\ell}|^{\alpha' k_{\ell} q}\sum_{i=1}^{n}  \frac{k_{i\mu}}{z- z_i} \sum_{j=1}^{n}  \frac{k_{j\nu}}{{\bar{z}} - {\bar{z}}_j} \, .
\label{Nzbarzqki}
\end{eqnarray}
and $\kappa_d$ is the gravitational coupling constant. The quantities $z_i$ are complex coordinates
parametrizing the insertion on the world-sheet of the vertex operators
associated to  the tachyon states. The coordinate~$z$, without index, is
associated to the massless closed string state. Finally,  the soft momentum of massless states is denoted by $q$. 

In principle $M_{\mu\nu}$ describes  also the emission of one anti-symmetric tensor  from a scattering  amplitude  with $n$ tachyons. However, this latter contribution vanishes because the world-sheet parity $\Omega$ leaves the vertex operators of the tachyon, dilaton, and graviton invariant, while changes the sign of the vertex operator of the Kalb-Ramond. 

The main aspect of these new soft theorems consists in finding an 
operator, $\hat{S}$, that acting on $n$-point amplitudes reproduces 
the soft behaviour of $n+1$-point amplitudes. The soft operator is 
determined by evaluating eq. (\ref{Nzbarzqki}) for small $q$. 
Eq. (\ref{Nzbarzqki}) is a sum of integrals on the complex plane that have 
been explicitly  computed in Ref.\cite{1502.05258}. Here we only quote the result:
\begin{eqnarray}
\label{totalexpre}
&& S_{\mu \nu} = 2\pi\Bigg\{  \sum_{i=1}^{n} k_{i\mu} k_{i \nu}  \Bigg[    \frac{(\alpha')^2}{2}
 \sum_{j \neq i} (k_j q) \log^2 |z_i - z_j| \nonumber\\
 && + \frac{1}{k_i q} \Big( 1 +\alpha'  \sum_{j \neq i} (k_j q)  
  \log |z_i - z_j| + \frac{(\alpha')^2}{2} 
\sum_{j;k \neq i}(k_j q) (k_k q)\nonumber\\
&&  \log|z_i -z_j|   
\log |z_i - z_k| \Big) \Bigg]+\sum_{i \neq j} \frac{k_{i\mu} k_{j\nu} + k_{i\nu} k_{j\mu}}{2}\nonumber\\
&&\times \Bigg[ - \alpha'  \log|z_i-z_{j}|
+   \frac{(\alpha' )^2}{2}  \Bigg( \sum_{k \neq i,j} (k_k q)   \Big( \log |z_k - z_{i}| \nonumber\\
&&\log |z_k - z_{j}| \Big)- \sum_{k \neq i} (k_k q)   \log|z_i - z_{j}| \log |z_k - z_{i}|
\nonumber\\
&&- \sum_{k \neq j} (k_k q) \log|z_i - z_{j}| \log |z_k - z_{j}| \Bigg) \Bigg]\Bigg\} + 
O(q^2) \ .
\end{eqnarray}
After a long but straightforward calculation the result of the integrations is rewritten 
in terms of the differential operators acting on the $n$-tachyon amplitude:
 \begin{eqnarray}
 &&\frac{M_{\mu \nu}}{\kappa_d}=
    \sum_{i=1}^{n} 
    \Bigg[ \frac{k_{i\mu} k_{i\nu}}{k_i q}  - i \frac{k_{i\nu} q^\rho L^{(i)}_{\mu \rho} }{2k_i q}  - i \frac{k_{i\mu} q^\rho L^{(i)}_{\nu \rho} }{2k_i q}\nonumber\\
    && -  \frac{q^{\rho} L_{i \,\mu \rho} q^{\sigma} L_{i \,\nu \sigma}  }{2k_i q}   + \left(  \frac{1}{2} \left(\eta_{\mu \nu}  q_{\sigma} - q_\mu \eta_{\nu \sigma}\right)      - \frac{k_{i \mu} q_\nu q_\sigma  }{ 2k_iq} \right)  \frac{\partial}{\partial k_{i \sigma}} \Bigg]\nonumber\\
    &&\times T_n(k_1,\dots k_n) 
 + O(q^2) \ .\label{1gra4tacbvv}
 \end{eqnarray}
  with $T_n$ the $n$ tachyon amplitude defined for example in Ref.\cite{1502.05258} 
and $L_i$ are the angular momentum operators given by:
\begin{eqnarray}
L_i^{\mu \rho} = i \left( k_{i}^{\mu} \frac{\partial}{\partial k_{i\rho} }-  k_{i}^{\rho} \frac{\partial}{\partial k_{i\mu}}   \right) \,.
\label{Jmurho}
\end{eqnarray}
The scattering of the dilaton, graviton and Kalb-Ramond  is selected by saturating $M_{\mu\nu}$ with the projectors:
\begin{eqnarray}
\mbox{Graviton}\, \, (g_{\mu \nu})  \,\,\, &\Longrightarrow& \epsilon^{\mu \nu}_g = \epsilon^{\nu \mu}_g  \,\,\, ; \,\,\, \eta_{\mu \nu} \epsilon^{\mu \nu}_g =0  \label{epsG} \\
 \mbox{Dilaton } \, (\phi)  \,\,\, &\Longrightarrow&   \epsilon^{\mu \nu}_d =  \eta^{\mu \nu} - q^{\mu} {\bar{q}}^{\nu}  - q^{\nu} {\bar{q}}^{\mu} \label{epsd} \\
 \mbox{Kalb-Ramond }(B_{\mu \nu} )  \,\,\,  &\Longrightarrow&   \epsilon^{\mu \nu}_B = - \epsilon^{\nu \mu}_B 
\label{epsB}
\end{eqnarray}
where ${\bar{q}}$ is a lightlike vector such that \mbox{$q \cdot {\bar{q}}=1$}.

In the case of a graviton, we can neglect  the last three terms in the squared bracket  of Eq.~(\ref{1gra4tacbvv}) and we get 
\begin{eqnarray}
&&\epsilon^{\mu \nu}_g \frac{M_{\mu \nu} (q; k_i )}{\kappa_d}  =   \epsilon^{\mu \nu}_g   \sum_{i=1}^{n} \Bigg[ \frac{k_{i\mu} k_{i\nu}}{k_i q} - i \frac{k_{i\nu} q^\rho L^{(i)}_{\mu \rho} }{2 k_i q} \nonumber\\
&&-  i \frac{k_{i\mu} q^\rho L^{(i)}_{\nu \rho} }{2 k_i q}  - \frac{1}{2} \frac{q_{\rho} L_i^{\mu \rho} q_{\sigma} )^{\nu \sigma}_i  }{k_i q} \Bigg]T_n (k_i ) +O(q^2)\ , 
\label{Mgravi}
\end{eqnarray}
which, of course, agrees with the soft theorem for the graviton derived in section 3  of  Ref.~\cite{BDDN}.

In the case of the dilaton
one gets instead:
\begin{eqnarray}
&& \epsilon^{\mu \nu}_d \frac{M_{\mu \nu} (q; k_i)}{\kappa_d}  = 
 \Bigg[ \! - \!\sum_{i=1}^{n} \frac{ m_i^2 \left( 1 + q^{\rho}  \frac{\partial}{\partial k_{i}^{\rho}}
+ \frac{1}{2} q^{\rho}   q^{\sigma} \frac{ \partial^2}{ \partial k_{i}^{\rho} \partial k_{ i}^{\sigma}   } \right)
}{k_i q} \nonumber\\
&&  +2  -    \sum_{i=1}^{n} k_{i}^{\rho}   \frac{\partial}{ \partial k_{i}^{\rho }}-   \sum_{i=1}^{n}  \Bigg(k_{i\mu} q_{\sigma} \frac{\partial^2}{\partial k_{i\mu} \partial k_{i\sigma}}  - \frac{1}{2} (k_i q) \nonumber\\ 
&&\times \frac{\partial^2}{\partial k_{i\mu} \partial k_{i\mu}} \Bigg)\Bigg]   
 T_n (k_1,\dots k_n ) +O(q^2)\ ,
\end{eqnarray}
where $m_i^2 = -\frac{4}{\alpha'}$ is the squared mass of the closed string tachyon. 
The dilaton contains terms $O( q^{-1} )$ when the other particles are massive, because the three-point amplitude involving a dilaton and two equal particles with mass $m$ is  proportional to $m^2$.

We have then studied the 
%infrared  
soft behaviour of amplitudes involving only massless states. In this case the 
analysis has been done up to the subleading order in  the soft expansion and 
the result is rather complicated but  can be written as a convolution integral, $M_{n+1}=S\ast M_{n}$, 
with  $M_n$ the amplitude of $n$ massless states in the closed bosonic string, 
given 
%for example 
in ref. \cite{1502.05258}, and 
\begin{eqnarray}
S=2\pi \epsilon_{q\mu} {\bar{\epsilon}}_{ q\nu} \Big( S_{q^{-1}}^{\mu\nu}+S_{q^0}^{\mu\nu}\Big)+O(q)
\end{eqnarray}
with 
\begin{eqnarray}
S_{q^{-1}}^{\mu\nu}= \sum_{i=1}^{n}  \frac{k_{i}^{\mu} k_{i}^{\nu}}{k_i q}
\end{eqnarray}
and
\begin{eqnarray}
&&S_{q^0}^{\mu\nu} =  \sqrt{\frac{\alpha'}{2}}\sum_{j \neq i} \left[ \frac{\sqrt{2\alpha'}k_{i}^{\nu} q^{\rho}}{k_i q} \log |z_i - z_j|\left( k_{i}^{\mu} k_{j\rho} - k_{i\rho} k_{j}^{ \mu} \right)\right. \nonumber\\
&&  -   
\left(
\frac{\theta_i (\epsilon_i q)}{z_i - z_j} \left( \frac{k_{j}^{\mu} k_{i}^{\nu}}{k_i q} -  \frac{k_{j}^{\mu} k_{j}^{\nu}}{k_j q} \right) +\mbox{c.c}
  \right)- \left(  \frac{ \theta_i \epsilon_{i}^{\mu} k_{j}^{\nu}}{z_i - z_j}+\frac{ {\bar{\theta}}_i  {\bar{\epsilon}}_{i}^{\nu} k_{j}^{\mu}}{{\bar{z}}_i - {\bar{z}}_j}\right)\nonumber \\
&&\left.+     \frac{k_j q}{k_i q}\Bigg(  \frac{   \theta_i \epsilon_i^{\mu} k_{i}^{\nu}}{z_i - z_j}+\frac{   {\bar{\theta}}_i {\bar{\epsilon}}_i^{\nu} k_{i}^{\mu}}{{\bar{z}}_i-\bar{z}_j }\Bigg)\right] - \!\! \sum_{i \neq j} \!\!\frac{ (\theta_j \epsilon_j q)(\theta_i \epsilon_i^{\mu})}{(z_i - z_j)^2} 
\nonumber \\
&&\times \left( \frac{k_j^{\nu}}{k_j q} - \frac{k_i^{\nu}}{k_i q} \right)   - \sum_{i \neq j} \!\!\frac{ ({\bar{\theta}}_j {\bar{\epsilon}}_j q) ({\bar{\theta}}_i {\bar{\epsilon}}_i^{\nu})}{({\bar{z}}_i - {\bar{z}}_j)^2} \Bigg( \frac{k_j^{\mu}}{k_j q} - \frac{k_i^{\mu}}{k_i q} \Bigg) \ . 
\end{eqnarray}
The polarizations of the massless states are conveniently  written in the form $\epsilon_{\mu\nu}=\theta\epsilon_\mu\bar{\theta}\bar{\epsilon}_\nu$, 
with ($\theta,\, \bar{\theta}$) Grassmanian variables  and c.c. denotes the complex conjugate.

If we use the polarization of a graviton, given in Eq.~(\ref{epsG}), we get the soft behavior for a graviton in agreement with the result of Ref.~\cite{BDDN}. In the case of the dilaton we get instead:
\begin{eqnarray}
\frac{M_{n+1}}{\kappa_d}= 
 \left[ 2 - \sum_{i=1}^{n}  k_{i\mu} \frac{\partial}{\partial k_{i\mu}}  \right] M_n 
+ O (q ) \, ,
\label{sofdilafin}
\end{eqnarray}
 in agreement with the result of Ref.~\cite{Ademollo:1975pf}. We have  
checked that the previous  soft behaviour is also obtained in the case of the 
superstring.
 
In order to define  a low energy theorem for the antisymmetric tensor, it is convenient to keep distinct the holomorphic, $k_i$, and anti-holomorphic, ${\bar{k}}_i$,  momentum  coming from the factorized structure of the vertices in closed string theory. 
 According to such a separation one gets:
\begin{eqnarray}
&&\frac{iM_{n+1} }{\kappa_d}
 =\epsilon_{q\, \mu\nu}^B \sum_{i=1}^{n}   \left[ \frac{( L_i - {\bar{L}}_i )^{\mu \nu}}{2} + \frac{k_i^{\nu} q_{\rho}( S_i - {\bar{S}}_i)^{\mu \rho}}{k_i q}  \right]M_n\Bigg|_{k=\bar{k}}\nonumber\\
 &&\label{Bmunu44}
\end{eqnarray}
with 
\begin{eqnarray}
S_i^{\mu\nu}=i\left( \epsilon_i^\mu\frac{\partial }{\partial \epsilon_{i\nu}} -\epsilon_i^\nu\frac{\partial }{\partial \epsilon_{i\mu}}\right) \, , \  
{\bar{S}}^{\mu\nu}_i=i\left( {\bar{\epsilon}}_i^\mu\frac{\partial }{\partial {\bar{\epsilon}}_{i\nu}} -{\bar{\epsilon}}_i^\nu\frac{\partial }{\partial {\bar{\epsilon}}_{i\mu}}\right)  \ ,
\end{eqnarray}
and $\bar{L}$ is the angular momentum operator written in terms of the anti-holomorphic momenta $\bar{k}$.   Eq. (\ref{Bmunu44}) reproduces the soft behavior of the antisymmetric tensor,  
but it is not a real soft theorem as in the case of the graviton and dilaton because, due to the separation of $k$ and ${\bar{k}}$,  the amplitude $M_n$ is not a physical amplitude.  

\section{Double soft limit of string amplitudes}
In this section we consider the color-ordered scattering amplitude, $M_{2g;ng}$, involving 
$(n+2)$ gauge fields living on the world-volume of a D$p$-brane 
of the bosonic string and we compute the leading double-soft behavior
when two contiguous  gluons become simultaneously soft. 

We denote with $(\epsilon_{q_1} , q_1)$ and  $(\epsilon_{q_2} , q_2)$  
the polarizations and momenta of two contiguous  gluons that eventually will become
soft and with $(\epsilon_i , k_i)$ the polarizations and 
momenta of the remaining gluons. 
The amplitude has been computed in detail in ref.\cite{1502.05258}, and here we give only the result written as a convolution integral, $M_{2g;ng}= M_{ng} * G_n$,  between the $n$ gluon amplitude and the  quantity $G_n$ which contains all the information about double soft behaviour of the gluons  
 \begin{eqnarray}
&& G_n =   2 \alpha' g_{p+1}^2  \int_{0}^{z_{n-1}} dw_1  \int_{0}^{w_1} dw_2 (w_1- w_2)^{2\alpha' q_1 q_2} 
 \nonumber\\
 &&\times
   \prod_{i=1}^{n} \prod_{a=1}^2\Bigg[(z_i - w_a)^{2\alpha' k_i q_a}
 {\rm e}^{\sqrt{2\alpha'}  \frac{\theta_i \epsilon_i q_a}{z_i - w_a}}\Bigg]\Bigg\{   \frac{ (\epsilon_{q_1} \epsilon_{q_2})}{(w_1 - w_2)^2}
  \nonumber \\
   &&     
 + 
  \left[ \sum_{i=1}^{n} \frac{\theta_i ( \epsilon_i  \epsilon_{q_1} )}{(z_i - w_1)^2} 
  - \sum_{i=1}^{n} \frac{\sqrt{2\alpha'} (k_i \epsilon_{q_1} )}{z_i - w_1}  
   + 
 \frac{\sqrt{2\alpha'} (\epsilon_{q_1} q_2 )}{w_1 - w_2}    \right] 
  \nonumber \\
 && \left.  \times  \left[ \sum_{j=1}^{n} \frac{\theta_j 
 ( \epsilon_j  \epsilon_{q_2} )}{(z_j - w_2)^2}  - \sum_{j=1}^{n}
  \frac{\sqrt{2\alpha'} (k_j \epsilon_{q_2} )}{z_j - w_2}       -
  \frac{\sqrt{2\alpha'} (\epsilon_{q_2} q_1 )}{w_1 - w_2}    \right]  \right\} \nonumber .
 \label{softfactorSbis}
 \end{eqnarray}
The latter integral has been computed in the limit of small momenta $q_1$, $q_2$
and the resulting expression turns out to be:
 \begin{eqnarray}
 &&G_n   =  
 \frac{g_{p+1}^{2} }{ q_1 q_2}  
  \Bigg\{
     \Bigg[- \frac{(\epsilon_{q_1} \epsilon_{q_2} )k_n (q_2 - q_1) + 
 q_1 q_2 }{2s_{n} } \nonumber\\
 &&  
 + \frac{ (\epsilon_{q_1} q_2 ) (\epsilon_{q_2} k_n) -  (\epsilon_{q_2} q_1 ) 
  (\epsilon_{q_1} k_n)}{  s_n } + \frac{(\epsilon_{q_1} k_n) (\epsilon_{q_2} k_n) (q_1 q_2)}{(k_n q_2) 
   s_n}
   \nonumber \\
  && + k_{n}\leftrightarrow k_{n-1}\Bigg]
  - \frac{(\epsilon_{q_1} k_{n-1}) ( \epsilon_{q_2} k_n) 
 (q_1 q_2) }{(k_{n-1} q_1 ) (k_n q_2)}
  \Bigg\}, \label{a}
 \end{eqnarray}
where  $s_\alpha=k_\alpha(q_1+q_2)+q_1q_2$  with $\alpha=n,n-1$.  
 Eq.(\ref{a}) is gauge invariant and  behaves as $\frac{1}{q_{1} q_2}$ in the double-soft limit,  i.e. when both $q_1$ and 
 $q_2$ simultaneously go to zero.

The double-soft 
%infrared 
behaviour of the $n+2$-point color ordered amplitude with two soft particles 
separated by a hard one is evaluated along the same lines of the contiguous case.
The resulting expression is again a convolution between the $n$ point gluon 
amplitude and the momenta dependent quantity:
\begin{eqnarray}
 G_{2g} &=&  g_{p+1}^2 \left[ \frac{ k_{n-2} \epsilon_{q_1}}{k_{n-2} q_1 } 
\left(  \frac{k_{n-1} \epsilon_{q_2}}{k_{n-1} q_2} -  \frac{k_{n} 
\epsilon_{q_2}}{k_{n} q_2} \right) +  \frac{k_{n-1} \epsilon_{q_1}}{k_{n-1} q_1}
 \frac{k_{n} \epsilon_{q_2}}{k_{n} q_2}  \right. \nonumber \\
   &-&  \left. \frac{ (\epsilon_{q_1} k_{n-1})  (\epsilon_{q_2} k_{n-1}) }{ 
k_{n-1}(q_1 + q_2) + q_1 q_2   } \left( \frac{1}{ k_{n-1} q_1} +
 \frac{1}{ k_{n-1} q_2}   \right)  \right]
\label{sidstesidste} 
\end{eqnarray}     
It is easy to check that the soft factor is gauge invariant up to terms of order
$q_{1,2}^0$ as in the case of two contiguous soft gluons.

\section{Concluding Remarks}
We have presented the results of 
%our works in 
Ref.\cite{1502.05258}, showing that bosonic string theory is a useful framework for computing low-energy properties of scattering amplitudes in a gauge covariant way and for any dimensions. The framework also allows to extend the results straightforwardly to higher orders in soft-momenta and to directly apply it to superstrings, which we plan to present in a future work.

\end{document}